\documentclass[prc,superscriptaddress,noshowpacs,unsortedaddress,twocolumn,showpacs,preprintnumbers,amsmath,amssymb]{revtex4-1}

\usepackage[dvipdfmx]{graphicx}
\usepackage{amsmath,amssymb,times}
\usepackage{color}
\usepackage{ulem}
\usepackage{bm}

\newcommand{\beq}{\begin{equation}}
\newcommand{\eeq}{\end{equation}}
\newcommand{\bea}{\begin{eqnarray}}
\newcommand{\eea}{\end{eqnarray}}

\begin{document}
\title{Neutron skin  $r_{\rm skin}^{48}$ determined from reaction cross section of proton+$^{48}$Ca scattering 
}

\author{Shingo~Tagami}
\affiliation{Department of Physics, Kyushu University, Fukuoka 819-0395, Japan}

\author{Maya~Takechi}
\affiliation{Niigata University, Niigata 950-2181, Japan}

\author{Jun~Matsui}
\affiliation{Department of Physics, Kyushu University, Fukuoka 819-0395, Japan}

\author{Tomotsugu~Wakasa}
\affiliation{Department of Physics, Kyushu University, Fukuoka 819-0395, Japan}

\author{Masanobu Yahiro}
\email[]{orion093g@gmail.com}
\affiliation{Department of Physics, Kyushu University, Fukuoka 819-0395, Japan}             

\date{\today}

\begin{abstract}
\noindent 
{\bf Background:}
Using the chiral  (Kyushu) $g$-matrix folding model with 
the densities calculated with Gongny-HFB (GHFB) with the angular momentum  projection (AMP), 
we determined the neutron skin $r_{\rm skin}^{208}=0.25$~fm from the central values of 
reaction cross sections $\sigma_{\rm R}({\rm EXP})$ of p+$^{208}$Pb scattering in $E_{\rm in}=40-81$~MeV. 
As for $^{48}$Ca, the high-resolution $E1$ polarizability experiment ($E1$pE) 
yields $r_{\rm skin}^{48}(E1{\rm pE}) =0.14-0.20~{\rm fm}$. Meanwhile,  
$\sigma_{\rm R}({\rm EXP})$ are available as a function of incident energy $E_{\rm in}$ 
for $p$+$^{48}$Ca scattering.  
 \\
{\bf Aim:} 
Our aim is to determine  $r_{\rm skin}^{48}$ from the central values of 
$\sigma_{\rm R}({\rm EXP})$  for $p$+$^{48}$Ca scattering by using the Kyushu $g$-matrix folding model. 
\\
{\bf Results:}  
As for $^{48}$Ca, we first determine neutron radius $r_n(E1{\rm pE})=3.56$~fm from the central value 0.17~fm 
of $r_{\rm skin}^{48}(E1{\rm pE})$ and the proton radius $r_p({\rm EXP})=3.385$~fm of electron scattering, 
and evaluate matter radius $r_m(E1{\rm pE})=3.485$~fm from the  $r_n(E1{\rm pE})$ and the $r_p({\rm EXP})$ 
of electron scattering. 
The Kyushu $g$-matrix folding model with the GHFB+AMP densities 
reproduces  $\sigma_{\rm R}({\rm EXP})$ in $23 \leq E_{\rm in} \leq 25.3$~MeV in one-$\sigma$ level, 
but slightly overestimates the central values of $\sigma_{\rm R}({\rm EXP})$ there.
In $23 \leq E_{\rm in} \leq 25.3$~MeV, the small deviation allows us to scale the  GHFB+AMP densities to 
the central values of  $r_p({\rm EXP})$ and  $r_n(E1{\rm pE})$. 
The reaction cross sections $\sigma_{\rm R}(E1{\rm pE})$ obtained with the scaled densities almost reproduce 
the central values of $\sigma_{\rm R}({\rm EXP})$ when $E_{\rm in}=23-25.3$~MeV, 
so that the $\sigma_{\rm R}({\rm AMP})$ obtained with the GHFB+AMP densities and the $\sigma_{\rm R}(E1{\rm pE})$ are 
in one $\sigma$ of $\sigma_{\rm R}({\rm EXP})$ there. 
In $E_{\rm in}=23-25.3$~MeV, we determine  the $r_{m}({\rm EXP})$ from the central values of 
$\sigma_{\rm R}({\rm EXP})$ and take the average for the $r_{m}({\rm EXP})$. The averaged value is 
$r_{m}({\rm EXP})=3.471$~fm. Eventually, we obtain $r_{\rm skin}^{48}({\rm EXP})=0.146$~fm from 
$r_{m}({\rm EXP})=3.471$~fm and $r_p({\rm EXP})=3.385$~fm. 
\end{abstract}

\maketitle


\section{Introduction and conclusion}
\label{Introduction}

{\it Background:}
As for neutron skin thickness $r_{\rm skin}=r_{\rm n}-r_{\rm p}$, 
Horowitz, Pollock and Souder proposed a direct measurement~\cite{Hor01a}.  
The measurement composes of parity-violating and elastic electron scattering. 
The neutron radius $r_{\rm n}$ is determined from the former, whereas
the proton radius $r_{\rm p}$ is from the latter. In fact, the $^{208}$Pb Radius EXperiment (PREX)~\cite{PREX05,Abrahamyan:2012gp} yields 
\bea
r_{\rm p}({\rm PREX})&=&5.45~{\rm fm},  
\label{Eq:direct constraint-Rp}
\\
r_{\rm n}({\rm PREX})&=&5.78^{+0.16}_{-0.18}~{\rm fm}, 
\label{Eq:direct constraint-Rn}
\\
r_{\rm skin}^{208}({\rm PREX})&=&0.33^{+0.16}_{-0.18}=0.15-0.49~{\rm fm} . 
\label{Eq:direct constraint-skin}
\eea
For $^{208}$Pb, the $r_{\rm p}({\rm PREX})$ and $r_{\rm n}({\rm PREX})$ are most reliable at the present stage. 
For $^{48}$Ca, the $^{48}$Ca Radius EXperiment (CREX) is ongoing at Jefferson Lab~\cite{PREX05}.

As an indirect measurement, the high-resolution $E1$ polarizability experiment ($E1$pE) yields 
\bea
r_{\rm skin}^{208}(E1{\rm pE}) &=&0.156^{+0.025}_{-0.021}=0.135-0.181~{\rm fm} 
\label{Eq:skin-Pb208-E1}
\eea
for $^{208}$Pb~\cite{Tamii:2011pv} 
\bea
r_{\rm skin}^{48}(E1{\rm pE}) &=&0.17 \pm 0.03^{+0.025}_{-0.021}=0.14-0.20~{\rm fm}~~~  
\label{Eq:skin-Ca48-E1}
\eea
for $^{48}$Ca~\cite{Birkhan:2016qkr}. 
As for $r_{\rm skin}$, the value for $^{48}$Ca is slightly larger than that for $^{208}$Pb.

The central value of reaction cross section $\sigma_{\rm R}$ is a standard observable
to determine matter radius  $r_{\rm m}$. 
One can  evaluate $r_{\rm skin}$ and $r_{\rm n}$ from the $r_{\rm m}$ and 
the $r_{\rm p}({\rm exp})$~\cite{Angeli:2013epw} of the electron scattering. 
Eventually, one can determine $r_{\rm skin}$ from the central value of  $\sigma_{\rm R}({\rm exp})$. 
The  $\sigma_{\rm R}({\rm exp})$ are available for $p$+$^{48}$Ca scattering~\cite{Carlson:1994fq}.

In the previous work~\cite{Tagami:2020bee}, we determined the neutron skin $r_{\rm skin}^{208}=0.25$~fm 
from the central values of $\sigma_{\rm R}({\rm EXP})$ of p+$^{208}$Pb scattering in $E_{\rm in}=40-81$~MeV, 
using the chiral  (Kyushu) $g$-matrix folding model with 
the densities calculated with Gongny-HFB (GHFB) with the angular momentum  projection (AMP), 

The $g$-matrix folding model is a standard way of obtaining microscopic optical potential 
for proton scattering and nucleus-nucleus scattering~\cite{Brieva-Rook,Amos,Satchler-1979,CEG, CEG07, Egashira:2014zda,Toyokawa:2014yma,Toyokawa:2015zxa,Toyokawa:2017pdd,Tagami:2019svt}. 
Applying the folding model with the Melbourne $g$-matrix~\cite{Amos} 
for  interaction cross sections $\sigma_{\rm I}$ for Ne isotopes and $\sigma_{\rm R}$ for Mg isotopes, 
we found that $^{31}$Ne is a deformed halo nucleus~\cite{Minomo:2011bb},   
and deduced the matter radii $r_{\rm m}$ for Ne isotopes~\cite{Sumi:2012fr} and  
for Mg isotopes~\cite{Watanabe:2014zea}. 

Kohno calculated the $g$ matrix  for the symmetric nuclear matter, 
using the Brueckner-Hartree-Fock method with chiral N$^{3}$LO 2NFs and NNLO 3NFs~\cite{Koh13}. 
He set $c_D=-2.5$ and $c_E=0.25$ so that  the energy per nucleon can  become minimum 
at $\rho = \rho_{0}$. 
Toyokawa {\it et al.} localized the non-local chiral  $g$ matrix into three-range Gaussian forms~\cite{Toyokawa:2017pdd}, using the localization method proposed 
by the Melbourne group~\cite{von-Geramb-1991,Amos-1994,Amos}. 
The resulting local  $g$ matrix is called  ``Kyushu  $g$-matrix''.

The  Kyushu $g$-matrix folding model is successful in reproducing $\sigma_{\rm R}$ 
 and differential cross sections  $d\sigma/d\Omega$ for $^4$He scattering 
 in $E_{\rm lab}=30 \sim 200$~MeV per nucleon~\cite{Toyokawa:2017pdd}. 
The success is true for proton scattering at $E_{\rm lab}=65$~MeV~\cite{Toyokawa:2014yma}. 

{\it Proton and neutron densities used in the folding model:}
In Ref.~\cite{Tagami:2019svt}, GHFB and GHFB+AMP reproduce the one-neutron separation energy $S_{1}$ and 
the two-neutron separation energy $S_{2}$ in $^{41-58}$Ca~\cite{HP:NuDat 2.7,Tarasov2018,Michimasa2018obr}. 
We found, with  $S_{1}$ and $S_{2}$, 
that $^{64}$Ca is an even-dripline nucleus and $^{59}$Ca is an odd-dripline nucleus. 
Our results are consistent with the data~\cite{HP:NuDat 2.7} in $^{40-58}$Ca for the binding energy $E_{\rm B}$. 
This means that the proton and neutron densities are good.

{\it Aim:} 
Our aim is to determine  a central value of $r_{\rm skin}^{48}$ from the central values of 
measured $\sigma_{\rm R}$  for $p$+$^{48}$Pb scattering, using the Kyushu $g$-matrix folding model  
with the GHFB+AMP densities. 

{\it Results:}
We first determine neutron radius $r_n(E1{\rm pE})=3.56$~fm from the central value 0.17~fm 
of $r_{\rm skin}^{48}(E1{\rm pE})$~\cite{Birkhan:2016qkr} and 
the proton radius $r_p({\rm EXP})=3.385$~fm of electron scattering, 
and evaluate matter radius $r_m(E1{\rm pE})=3.485$~fm from the  $r_n(E1{\rm pE})$ and the $r_p({\rm EXP})$. 

The Kyushu $g$-matrix folding model with the GHFB+AMP densities 
reproduces  $\sigma_{\rm R}({\rm EXP})$ in $23 \leq E_{\rm in} \leq 25.3$~MeV in one-$\sigma$ level. 
In $23 \leq E_{\rm in} \leq 25.3$~MeV, the small deviation allows us to scale the  GHFB+AMP densities to 
the central values $r_p({\rm EXP})$ and  $r_n(E1{\rm pE})$. 
The Kyushu $g$-matrix folding model with the scaled densities  almost reproduces the central values 
of $\sigma_{\rm R}({\rm EXP})$ when $E_{\rm in}=23-25.3$~MeV, so that both the $\sigma_{\rm R}({\rm AMP})$ 
obtained with the GHFB+AMP densities and the $\sigma_{\rm R}(E1{\rm pE})$ 
calculated with the scaled densities 
are in one $\sigma$ of $\sigma_{\rm R}(\rm EXP)$ there. 
In $E_{\rm in}=23-25.3$~MeV, we determine  the $r_{m}({\rm EXP})$ 
from the central values of $\sigma_{\rm R}({\rm EXP})$ 
by using $\sigma_{\rm R}({\rm EXP})=c r_{m}^2({\rm EXP})$ with $c=\sigma_{\rm R}(E1{\rm pE})/r_{\rm m}(E1{\rm pE})^2$, and take the average for the $r_{m}({\rm EXP})$.
In $E_{\rm in}=23-25.3$~MeV, we determine  the $r_{m}({\rm EXP})$ from the central values of $\sigma_{\rm R}({\rm EXP})$ and take the average for the $r_{m}({\rm EXP})$. The averaged value $r_{m}({\rm EXP})=3.471$~fm leads to 
$r_{\rm skin}^{48}({\rm EXP})=0.146$~fm.

{\it Conclusion:} 
Our conclusion is that the central value  of $r_{\rm skin}^{48}({\rm EXP)}$ is 0.146~fm. 
This result is consistent with  $r_{\rm skin}^{48}(E1{\rm pE})$ of Eq. \eqref{Eq:skin-Ca48-E1}.

\section{Model}
\label{Sec-Framework}

Our model is the Kyushu $g$-matrix  folding model~\cite{Toyokawa:2017pdd} 
with densities calculated with GHFB+AMP~\cite{Tagami:2019svt}.  
The folding model itself is clearly shown in Ref.~~\cite{Egashira:2014zda}. 
The Kyushu $g$-matrix is constructed from chiral interaction with the cutoff 550~MeV.

\section{Results}
\label{Results} 

\begin{figure}[htbp]
\begin{center}
 \includegraphics[width=0.5\textwidth,clip]{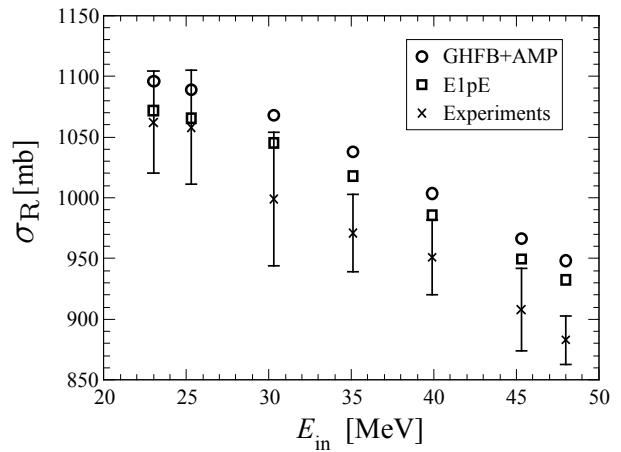}
 \caption{ 
 $E_{\rm in}$ dependence of reaction cross sections $\sigma_{\rm R}$ 
 for $p$+$^{48}$Ca scattering. 
 Circles denote results of the  GHFB+AMP densities, and 
 squares correspond to the folding model with the scaled densities.  
 The data (crosses) are taken from Refs.~\cite{Carlson:1994fq}. 
   }
 \label{Fig-RXsec-p+Ca48}
\end{center}
\end{figure}

Figure~ \ref{Fig-RXsec-p+Ca48} shows  reaction cross sections $\sigma_{\rm R}$ 
as a function of incident energy  $E_{\rm in}$ for $p$+$^{48}$Ca scattering. 
In one-$\sigma$ level, the Kyushu $g$-matrix folding model with the GHFB+AMP densities (circles) 
 reproduces  measured reaction cross sections $\sigma_{\rm R}({\rm EXP})$~\cite{Carlson:1994fq} 
in $23 \leq E_{\rm in} \leq 25.3$~MeV. 
Our results show that our framework is reliable there. 
This allows us to  scale the proton and neutron densities so that the proton and neutron radii, 
$r_p({\rm AMP})$ and $r_n({\rm AMP})$, calculated with GHFB+AMP may agree with 
the central values of $r_{\rm p}({\rm EXP})$ and $r_{\rm n}(E1{\rm pE})$. 
For $E_{\rm in}=23, 25.3$~MeV, the  Kyushu $g$-matrix folding model with the scaled densities (squares)  slightly overestimates the central values of $\sigma_{\rm R}({\rm EXP})$. 
In $E_{\rm in}=23, 25.3$~MeV, we determine  the $r_{m}({\rm EXP})$ 
from the central values of $\sigma_{\rm R}({\rm EXP})$ 
and take the average for the $r_{m}({\rm EXP})$; see {\it results} of Sec.~\ref{Introduction}
for how to determine $r_{m}({\rm EXP})$ from the central values of $\sigma_{\rm R}({\rm EXP})$.
The averaged value is $r_{m}({\rm EXP})=3.471$~fm. 
Using  the $r_{m}({\rm EXP})$ and $r_{\rm p}({\rm EXP})=3.385~{\rm fm}$, we can get 
$r_{\rm skin}^{48}=0.146$~fm.  

\noindent
\appendix

\noindent
\begin{acknowledgments}
We would like to thank Dr. Toyokawa for providing his code. 
\end{acknowledgments}



\bibliographystyle{prsty}

\end{document}